# Hydrostatic pressure study of paramagnetic-ferromagnetic phase transition in (Ga,Mn)As

M. Gryglas – Borysiewicz,[1] A. Kwiatkowski,[1] M. Baj,[1] D. Wasik,[1] J. Przybytek[1] and J. Sadowski[2]

[1]*Institute of Experimental Physics, Faculty of Physics, University of Warsaw, Hoża 69, 00-681 Warsaw, Poland*
*2Institute of Physics, PAS, Al. Lotników 32/46, 02-668 Warsaw, Poland and MAX-Lab, Lund University, 221 00 Lund, Sweden*

The effect of hydrostatic pressure on the paramagnetic – ferromagnetic phase transition has been studied in (Ga,Mn)As. The variation of the Curie temperature ($T_C$) with pressure was monitored by two transport methods: (1) – measurement of zero field resistivity versus temperature $\rho(T)$, (2) – dependence on temperature of the Hall voltage hysteresis loop. Two specimens of different resistivity characteristics were examined. The measured pressure-induced changes of $T_C$ were relatively small (of the order of 1K/GPa) for both samples, however they were opposite for the two.

(Ga,Mn)As is one of the most intensively investigated diluted magnetic semiconductors during last decades. The understanding of physical phenomena governing its magnetic properties is crucial for increasing Curie temperature ($T_C$) and thus for possible application of this material in spintronic devices. The origin of ferromagnetism in (Ga,Mn)As was quantitatively explained within the *p-d* Zener model assuming magnetic interaction between the localized magnetic moments of $Mn^{2+}$ ions mediated by holes in the valence band[1-3]. This model, in the case of semiconductors, where the carrier density is smaller than the magnetic ion concentration is equivalent to the Ruderman-Kittel-Kasuya-Yosida (RKKY) approach employed in the diluted magnetic metals[4]. Within this picture the ferromagnetic ordering temperature, $T_C$ depends in particular on the local *p-d* exchange interaction and a free hole concentration. It was demonstrated that indeed an increase of the hole concentration in a field effect transistor structure led to an enhancement of the ferromagnetic state[5]. On the other hand it was found that the exchange energy scales with the lattice constant as, $N_0\beta \sim a_0^{-3}$,[1] and therefore an external hydrostatic pressure could influence the exchange coupling. Although the studies of (In,Mn)Sb diluted magnetic semiconductor under hydrostatic pressure provided an evidence for an increase in carrier-mediated magnetic coupling[6,7], giving rise to higher Curie temperature, the effect of hydrostatic pressure on (Ga,Mn)As semiconductor is not as clear.[6] Therefore additional study was performed in order to clarify the role of external hydrostatic pressure in the paramagnetic-ferromagnetic phase transition in (Ga,Mn)As.

p-type $Ga_{1-x}Mn_xAs$ layers were grown by molecular beam epitaxy (MBE) on (100) GaAs substrate. In our studies two different samples were used: A777 and A963. The former sample had 20 nm thick layer of (Ga,Mn)As with Mn content $x$ = 7%. After the MBE growth this sample was capped with amorphous As and annealed in the MBE growth chamber at the temperature of 210 ºC (controlled by the IR pyrometer) for two hours (see Ref. 8 for details). The Curie temperature determined from SQUID magnetometry was close to 85 K (Fig. 1, open symbols). The latter sample had a (Ga,Mn)As layer of 50 nm and $x$ = 6%. This sample was not annealed after the MBE growth. The Curie temperature for A963 sample was approximately 50 K (Fig. 1, solid symbols). Since the amount of the magnetic material of our samples was very small, the SQUID signal was also very small. Precise measurements reveal also the presence of a very small high-temperature component in the magnetization. It was checked that this contribution did not originate from (Ga,Mn)As layer as it was present even after the layer was removed.

The effect of pressure on the paramagnetic-ferromagnetic phase transition in (Ga,Mn)As was studied by transport measurements under hydrostatic pressure up to 1 GPa at temperatures range 20 – 300 K. Standard Hall bars were prepared lithographically and six ohmic contacts were made with indium. The samples were placed in a beryllium-copper self-clamping high pressure cell. A mixture of kerosene and oil was used as a pressure transmitting medium.

To monitor pressure-induced changes of Curie temperature we adopted two procedures:

1. the study of zero-field resistivity $\rho(T)$; the position of maximum in $d\rho/dT$ was reported to coincide with the Curie temperature [9];
2. Hall resistivity measurements in a magnetic field perpendicular to the sample surface. In ferromagnetic phase the Hall resistivity is a sum of the ordinary Hall effect due to the Lorentz force and the anomalous Hall effect originating from the asymmetric scattering in the presence of magnetization[10]:

$$R_{Hall} = \frac{R_0 B}{d} + \frac{R_S \mu_0 M}{d} \qquad (1)$$

where, $R_0$ is the ordinary Hall coefficient, $B$ and $M$ the magnetic field and the magnetization perpendicular to the layer respectively, $R_S$ the anomalous Hall coefficient and $d$ the thickness of the (Ga,Mn)As layer. For $T < T_C$ hysteresis loop in $M$ and thus in $R_{Hall}(B)$ is observed. The studies of the evolution of the hysteresis provide the method to monitor $T_C$.

The results of the measurements of zero-field resistance versus temperature performed at ambient pressure are shown in Fig. 2. The behavior of $\rho(T)$ of the samples is quite different: A777 has rather metallic character, while A963 is clearly of semiconducting type. Characteristic rounded cusps are observed at about 100 K and 50 K for the sample A777 and A963, respectively, which are believed to be related to ferromagnetic – paramagnetic phase transition[11]. Figure 3 shows the evolution of the resistivity $\rho(T)$ dependence upon pressure. A clear shift towards lower temperatures is observed. At the same time the value of the resistivity increased by about 7%. Figure 4 shows temperature-derivative of resistivity $d\rho/dT$, calculated numerically and presented close to maxima, which can be identified with $T_C$ [9]. For both samples the maxima shift under pressure towards higher temperatures, although the shift is relatively small.

The results of the second method are shown in Fig. 5, which presents hysteresis loops of Hall voltage, measured for $T < T_C$. Remarkably, the shapes of hysteresis loops are very different for both samples, which indicates different magnetic easy axis for the samples studied (for A777 - in the sample plane, for A963 - perpendicular to plane). This is consistent with observations that for samples with lower hole concentrations easy axis can be perpendicular to plane[12]. Figure 6 presents the measured dependence of the width of the hysteresis loop on temperature for a few series of experiments made at various pressures for sample A963. One can see that hysteresis appears at temperatures close to 50 K. The results obtained for nominally the same values of pressure but during different experimental cycles differ slightly, which will be addressed later, but an overall pressure-induced decrease of $T_C$ measured with this method is clearly visible.

The summary of our experimental results is shown in Fig. 7. Referring to the absolute value of $T_C$, for sample A777 both methods give consistent values of $T_C$ which agree very well with the SQUID data of Fig. 1. For the sample A963 the two methods are inconsistent – the values of $T_C$ differ by more than 10 K. The results obtained from the hysteresis analysis agree with the SQUID data while the others do not. Most probably this is due to a diverging background in $\rho(T)$ dependence observed at low temperatures for A963 sample which shifts the maximum of $d\rho/dT$ towards lower temperatures. This also suggests that for samples that show strongly temperature-dependent semiconducting-type $\rho(T)$, the method proposed in Ref. 9 cannot be reliably applied. Referring to the pressure trends observed, for the A777 the application of hydrostatic pressure leads to the increase of Curie temperature. Both methods used reveal the similar trend, with a pressure coefficient of $dT_C/dp$ = + (2-3) K/GPa. This value is much higher (appr. 4 times) than the one found for (In,Mn)Sb[7]. However, for the sample A963 measurements of the hysteresis loop revealed an opposite pressure-induced shift ($dT_C/dp$ = — 1.0±0.2 K/GPa). As mentioned above, the second method is not credible enough to draw any conclusions, due to high impact of the diverging background. To sum up, our experimental results showed that pressure induced variation of Curie temperature is small and strongly depends on the metallic / semiconductor character of a sample.

Within RKKY interaction the temperature of ferromagnetic-paramagnetic phase transition is given by[1]:

$$T_C = x_{eff} S(S+1) \frac{(N_0\beta)^2}{N_0} \frac{A_F m^* k_F}{12 k_B \pi^2 \hbar^2} - T_{AF} \qquad (2)$$

where $x_{eff}$ is an effective manganese concentration, $S$ is the spin of Mn ion (=5/2), $N_0$ is a concentration of the cation sites, $N_0\beta$ is the *p-d* exchange energy, $k_F$ is the Fermi wavevector, $m^*$ is a valence band DOS effective mass, $k_B$ is the Boltzmann constant, $A_F$ is a constant describing carrier-carrier interactions, and $T_{AF}$ is a parameter describing antiferromagnetic superexchange interaction – for (Ga,Mn)As we put $T_{AF} \approx 0$.

External hydrostatic pressure can influence the following parameters from the Eq. (2):

1. Lattice cell volume and thus $N_0$. In our estimations the pressure induced change of $N_0$ was calculated according to the compressibility coefficients for GaAs[13],
2. Fermi vector, $k_F$. It is proportional to $p^{1/3}$ (where $p$ stands for the hole concentration). We assumed that $p$ changes due to volume changes only.
3. Exchange energy. According to Ref. 1 it should vary with a lattice constant as $N_0\beta \sim a_0^{-3}$ (i.e. $\beta$ can be treated as pressure-independent).
4. Heavy and light hole effective masses. To our knowledge it is not completely clear how these parameters change with pressure. However we adopted $\frac{1}{m_{lh}^*}\frac{dm_{lh}^*}{dp} = 0.074\ \mathrm{GPa}^{-1}$ and $\frac{1}{m_{hh}^*}\frac{dm_{hh}^*}{dp} = -0.001\ \mathrm{GPa}^{-1}$ after Ref. 14.

The pressure dependence of $T_C$ predicted by Eq. (2) is given by the solid lines in Fig. 7. The model reasonably well describes the pressure variation of $T_C$ for the sample A777 but is inconsistent with the experimental data obtained for the sample A963 from the analysis of the hysteresis loop. We believe that this striking difference in pressure shift of $T_C$ for both samples is related to different sample properties. Both samples have substantially different Curie temperatures and different easy axes, but what seems the most important for the understanding of the pressure shifts of their $T_C$, is the qualitatively different $\rho(T)$ dependences. Sample A777 is semi-metalic, while A963 is semiconducting. This means that in A963 the degree of localization of holes is much higher. According to the generalized alloy theory[15] the difference in the localization of holes may lead to both: different apparent values and different pressure dependences of *p-d* exchange energy $N_0\beta^{(app)}$ (and thus the Curie temperature). In particular, for samples close to metal-insulator transition, a decrease in the $T_C$ is expected when the localization is increased. To discuss the localization of Mn acceptor we recall experimental studies[16,17], showing that for Mn acceptor in Ga compounds (GaN, GaP, GaAs, GaSb) the smaller is the distance between the neighboring atoms, the more localized is the Mn acceptor. Under hydrostatic pressure we expect a similar trend – although the anion remains the same, the interatomic distances are reduced and in consequence localization increases. Such an interpretation is supported by the experimental observation of a strong increase of sample resistivity as a function of pressure (of the order of 30% for the pressure range used for A963, in contrast to 7% for A777).

There are two points that should be mentioned. Our experiments reveal some pressure-induced irreversible changes of sample properties (Fig. 7). Such observations of sample degradation are common in the pressure studies, when the lattice mismatch between the subsequent layers is increased by pressure (this is the case for (Ga,Mn)As on GaAs) and often they impede to make credible conclusions. However, due to tedious, several measurement cycles, the overall pressure-induced trend is clearly demonstrated. Another point is that in the model discussed the free hole concentration was assumed to change due to volume changes only. This does not necessarily have to be true. If in contrast one assumes that the observed pressure changes of the resistivity are entirely due to hole concentration (which is unrealistic), Eq. (2) would predict the decrease of $T_C$ for both samples with the rate of a few K per 1GPa. This means that there is a need of complementary studies of the influence of pressure on the free hole concentration in GaMnAs.

In summary, we have observed that in two (Ga,Mn)As samples differing in the degree of hole localization the pressure-induced variation of the Curie temperature was opposite. This finding is explained by a different dependence of $T_C$ upon increasing *p-d* hybridization depending on whether the sample is far or close to the metal-to-insulator transition[15].

We would like to thank Professor T. Dietl and Doctor A.Wołoś for valuable discussions.
J.S. acknowledges support from the FunDMS Advanced Grant of the European Research Council within the "Ideas" 7th Framework Programme of the EC.

[1]T. Dietl, H. Ohno, and F. Matsukura, Phys. Rev. B **63**, 195205 (2001).
[2]T. Jungwirth, J. König, J. Sinova, J. Kučera, and A. H. MacDonald, Phys. Rev. B **66**, 012402 (2002).
[3]T. Jungwirth, J. Sinova, J. Mašek, J. Kučera, and A. H. MacDonald, Rev. Mod. Phys. **78**, 809 (2006).
[4]T. Dietl, Semicond. Sci. Technol. **17**, 377 (2002).
[5]H. Ohno, D. Chiba, F. Matsukura, T. Omiya, E. Abe, T. Dietl, Y. Ohno, and K. Ohtani, Nature **408**, 944 (2000).
[6]M. Csontos, G. Mihály, B. Jankó, T. Wojtowicz, W. L. Lim, X. Liu, and J. K. Furdyna, phys. stat. sol. (c) **1**, 3571 (2004).
[7]M. Csontos, G. Mihály, B. Jankó, T. Wojtowicz, X. Liu, and J. K. Furdyna, Nature Materials **4**, 447 (2005).
[8]M. Adell, L. Ilver, J. Kanski, V. Stanciu, P. Svedlindh **,** J. Sadowski, J. Z. Domagala, F. Terki, C. Hernandez, S. Charar, Appl. Phys. Lett. **86**, 112501 (2005).
[9]V. Novák, K. Olejnik, J. Wunderlich, M. Cukr, K. Výborný, A. W. Rushforth, K. W. Edmonds, R. P. Campion, B. L. Gallagher, J. Sinova, and T. Jungwirth, Phys. Rev. Lett. **101**, 077201 (2008).
[10]C. M. Hurd, *The Hall Effect in Metals and Alloys,* Plenum Press (1972).
[11]F. Matsukura, H. Ohno, A. Shen, and Y. Sugawara, Phys. Rev. B **57**, R2037 (1998).
[12]M. Sawicki, F. Matsukura, T. Dietl, G. M. Schott, C. Ruester, G. Schmidt, L.W. Molenkamp and G. Karczewski, J.Supercond. **16**, 7 (2003)
[13]O. Madelung, *Semiconductors: Data Handbook,* Springer (2004).
[14]S. Adachi, *GaAs and Related Materials: Bulk Semiconducting and Superlattice,* World Scientific (1994).
[15]T. Dietl, Phys. Rev. B **77**, 085208 (2008).
[16]A. Wołoś, M. Piersa, G. Strzelecka, K. P. Korona, A. Hruban and M. Kaminska, phys. stat. sol. (C) **6**, 2769 (2009)
[17]B. Stepanek, P. Hubik, J. J. Mares, J. Kristofik, V. Sestakova, L. Pekarek and J Sestak, Semicond. Sci. Technol. **9**, 1138 (1994)

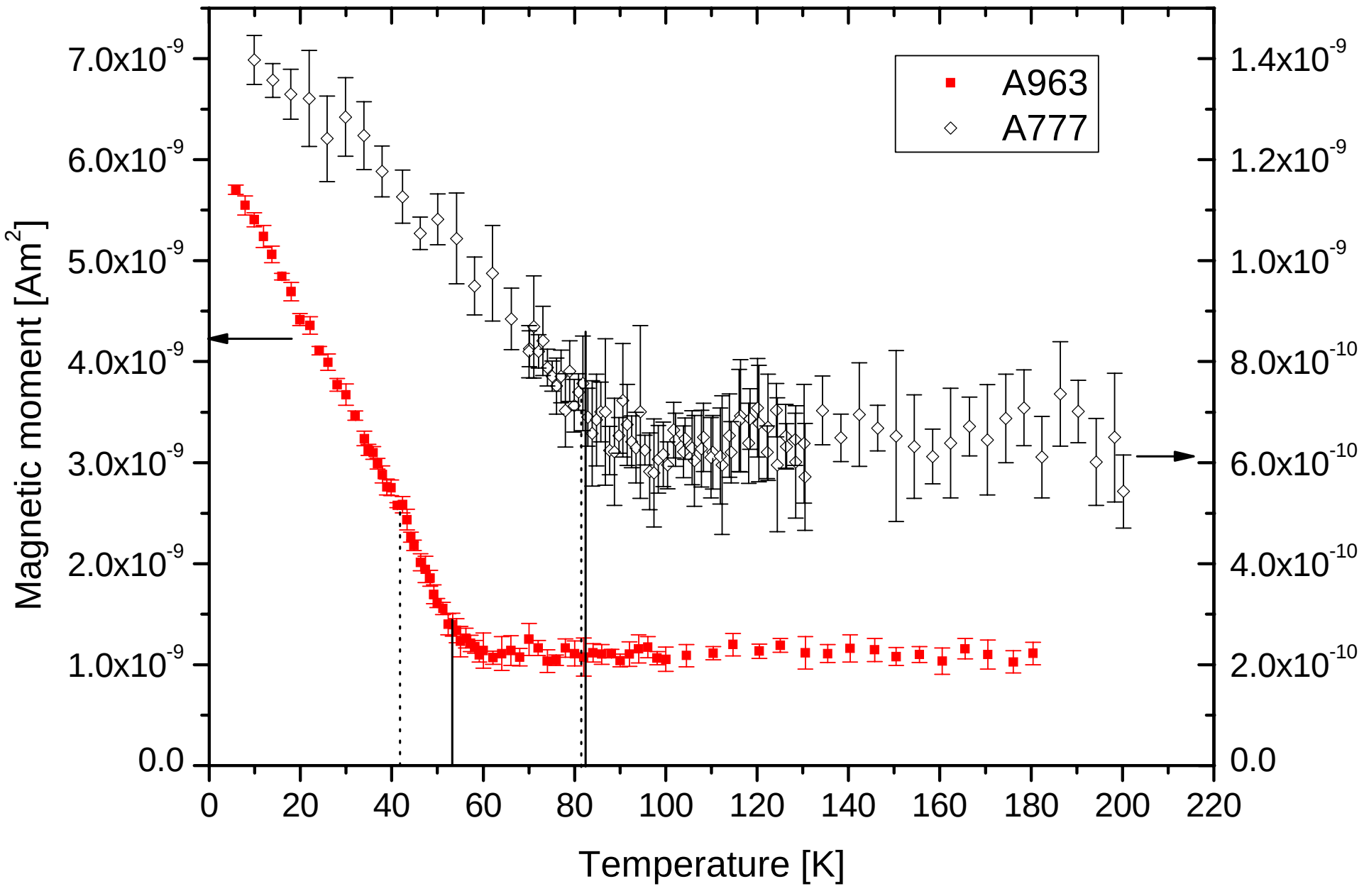

FIG. 1. Temperature dependence of magnetic moments measured by SQUID at $B$=2mT at ambient pressure for both samples (solid symbols – A963, open symbols – A777). The dotted and solid vertical lines depict the values of $T_C$ extracted from the maximum of $d\rho/dT$ and from the Hall voltage hysteresis, respectively (see text).

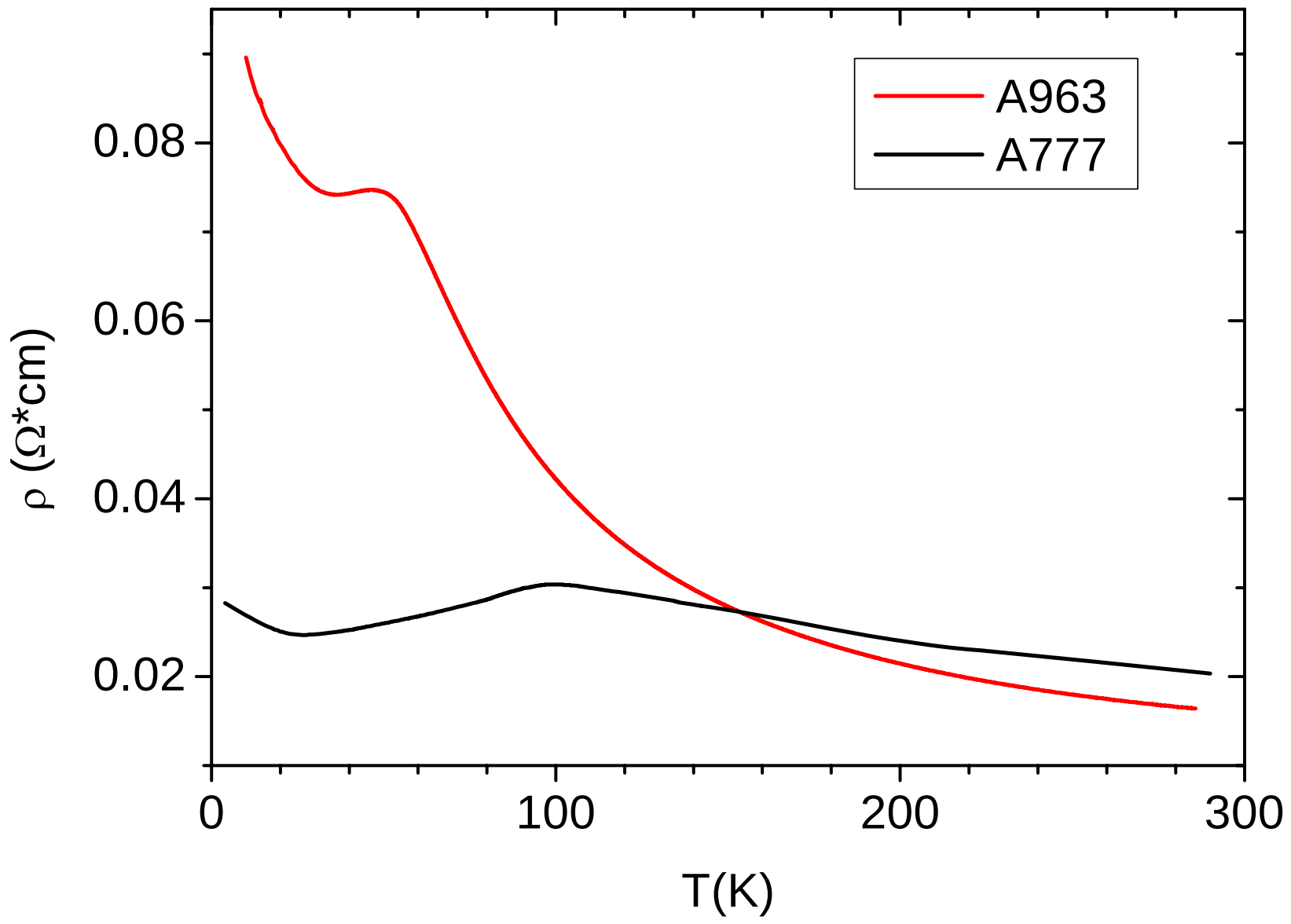

FIG. 2. Resistivity as a function of temperature $\rho(T)$ measured for the two samples at ambient pressure, before pressure cycles.

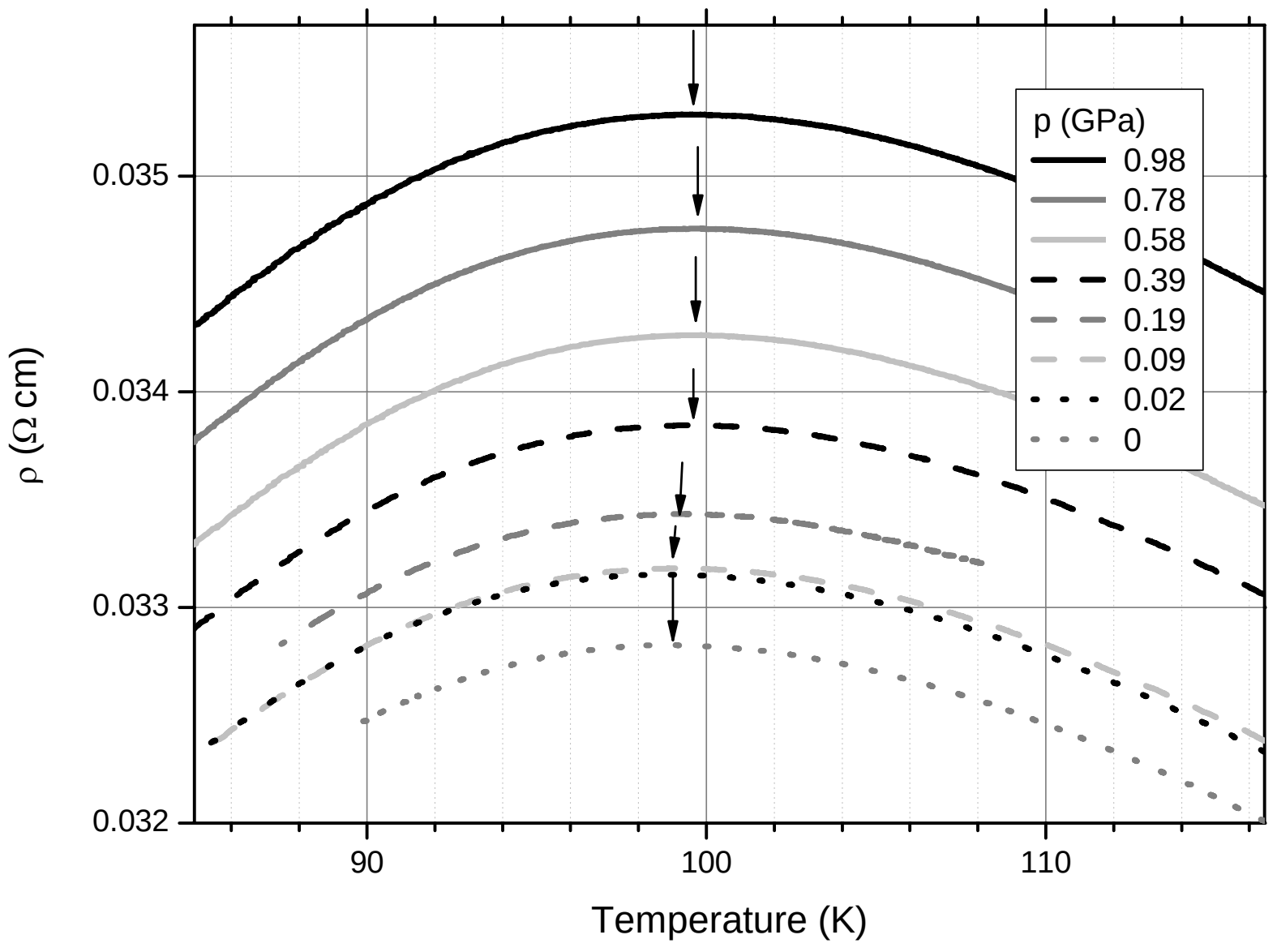

FIG.3. Resistivity evolution as a function of pressure for the A777 sample. Arrows show the evolution of the position and magnitude of the maximum in $\rho$(T).

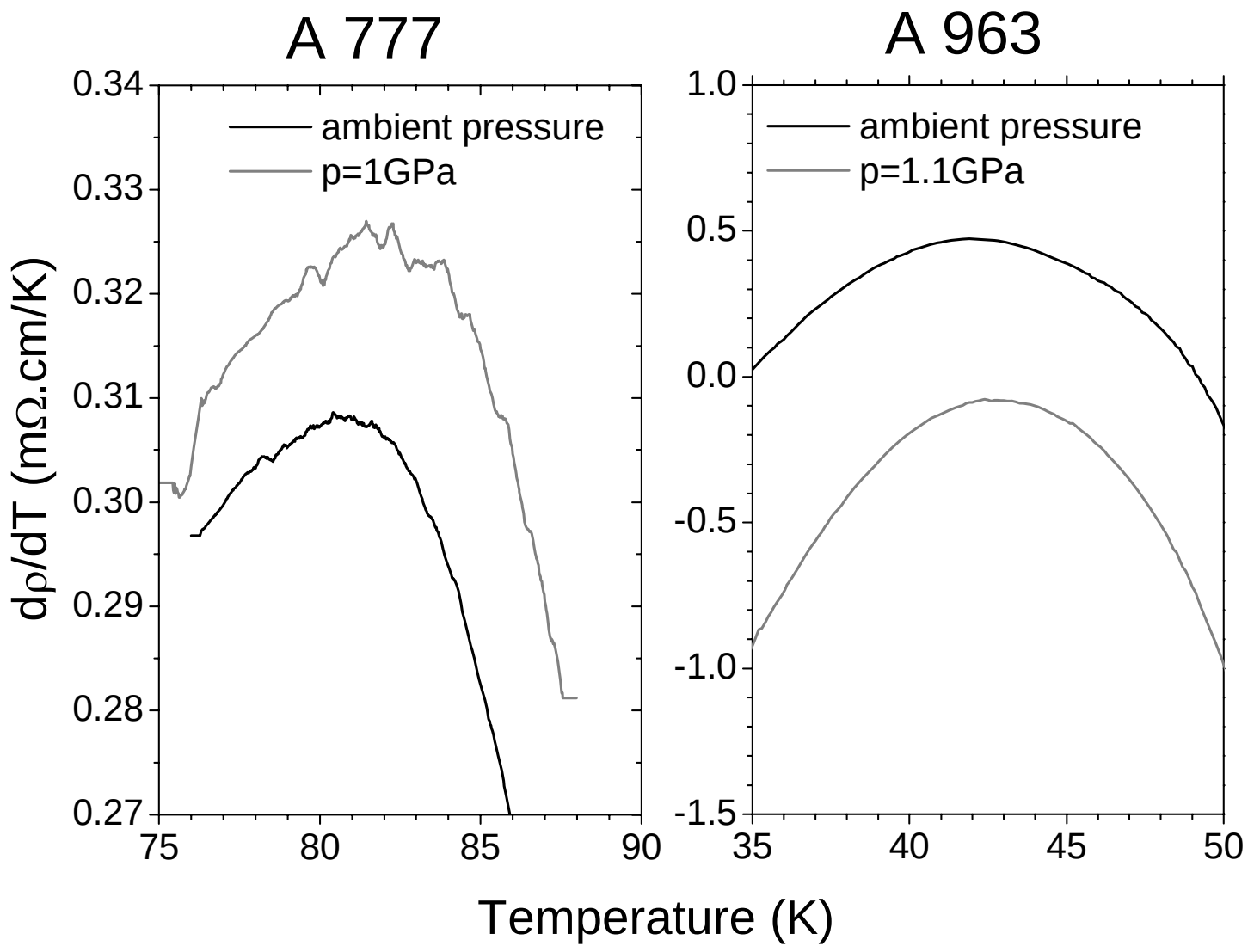

FIG. 4. Temperature derivative of the resistivity, *dρ/dT* calculated for both samples from the experimental data measured at ambient pressure and high pressure.

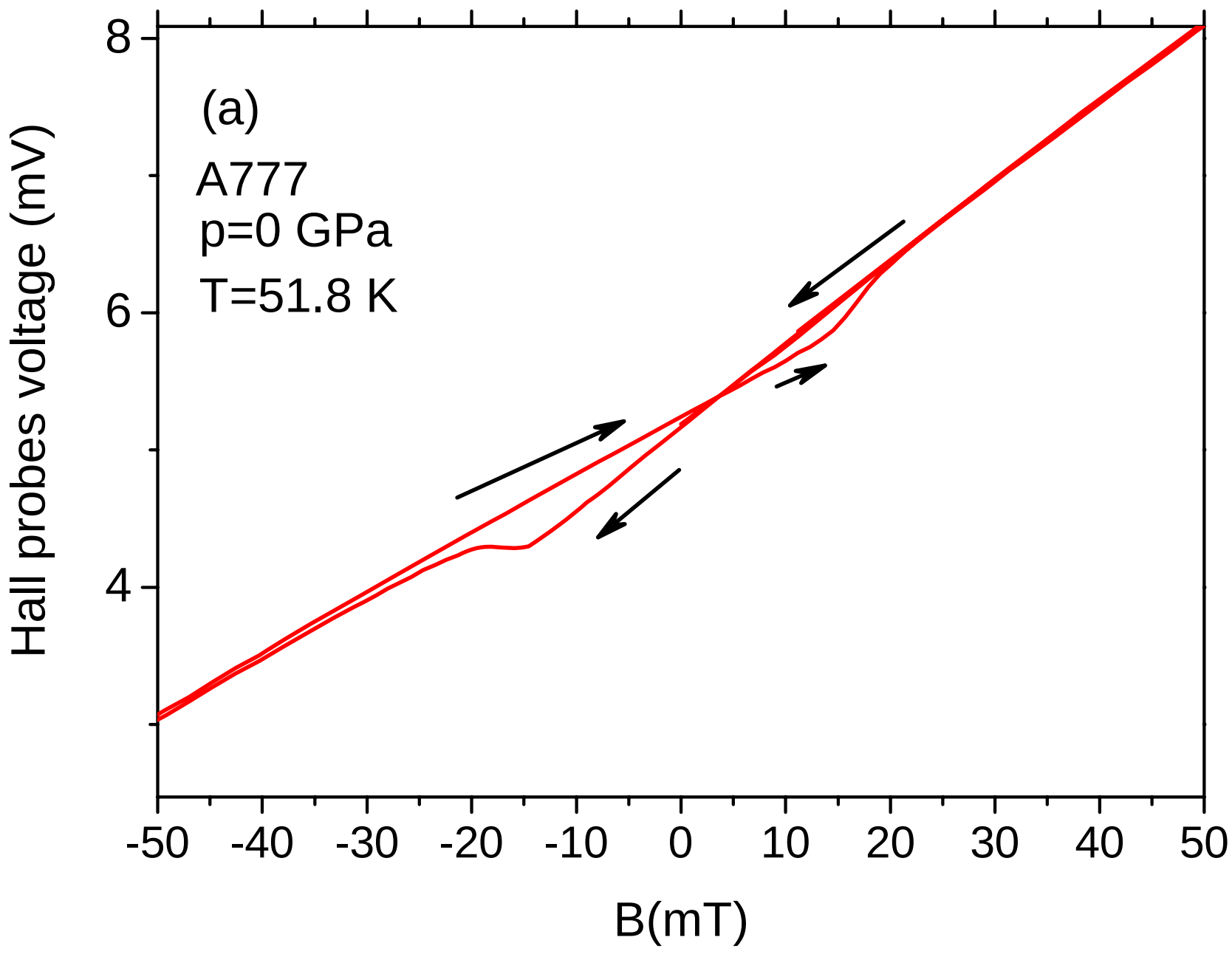

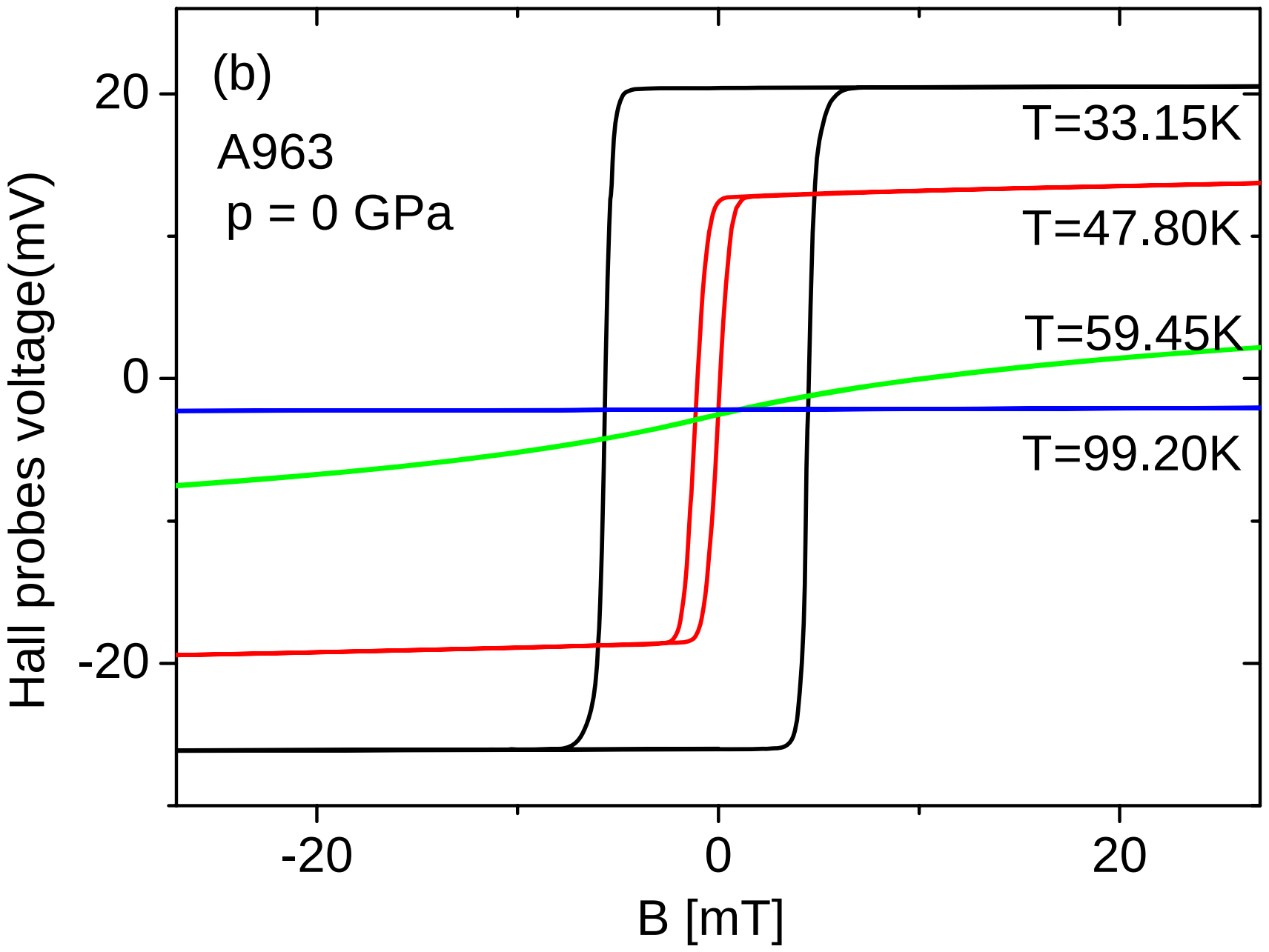

FIG. 5. Runs of the Hall voltage versus magnetic field revealing hysteresis loops for both our samples: (a) – A777, (b) – A963. Magnetic field was perpendicular to the sample plane. For the sample A963 the shape of the hysteresis loop indicates that an easy axis was perpendicular to the sample plane.

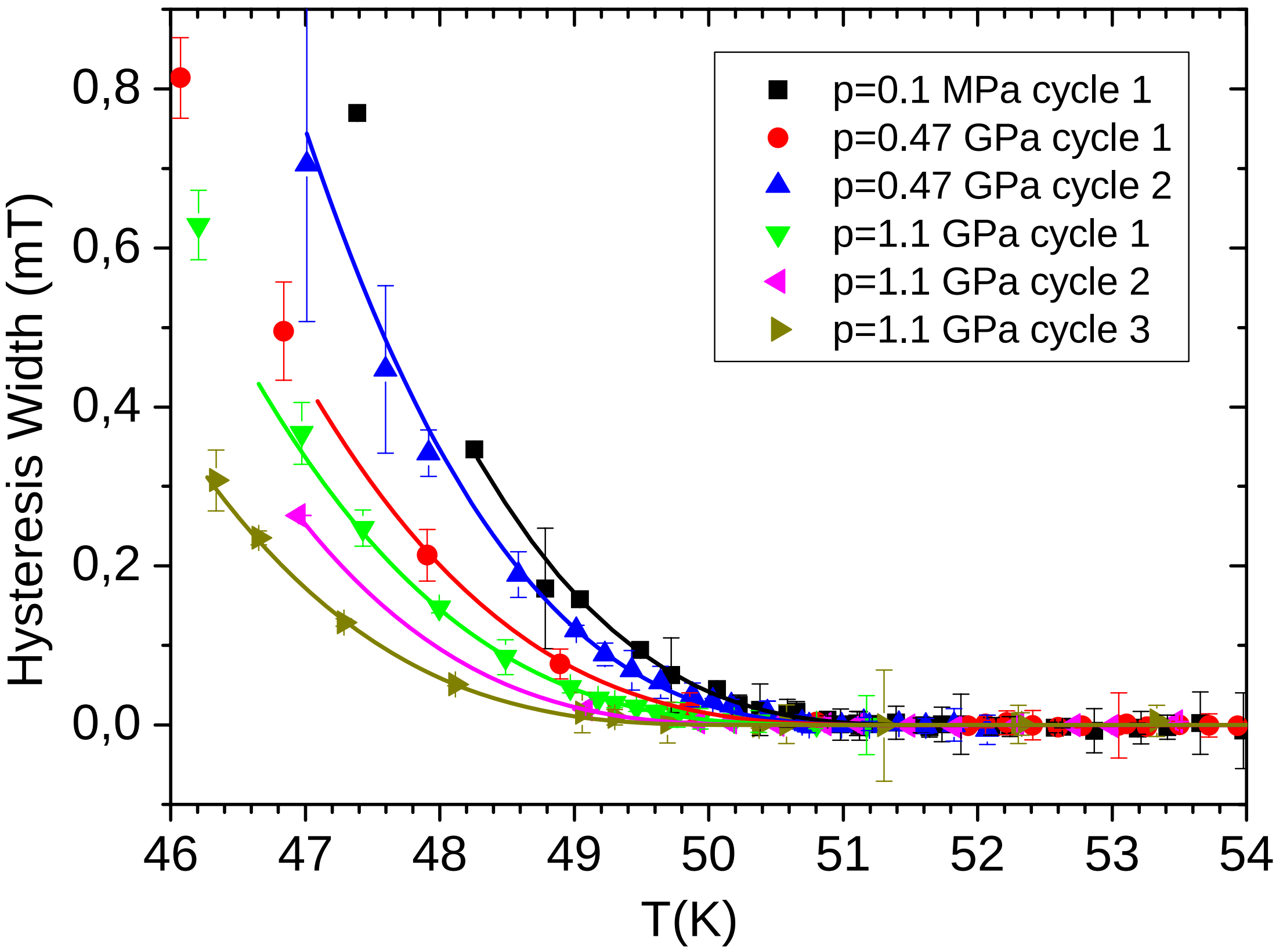

FIG. 6. Temperature dependence of the width of the hysteresis loop measured for the sample A963 in a series of experiments made at various pressures.

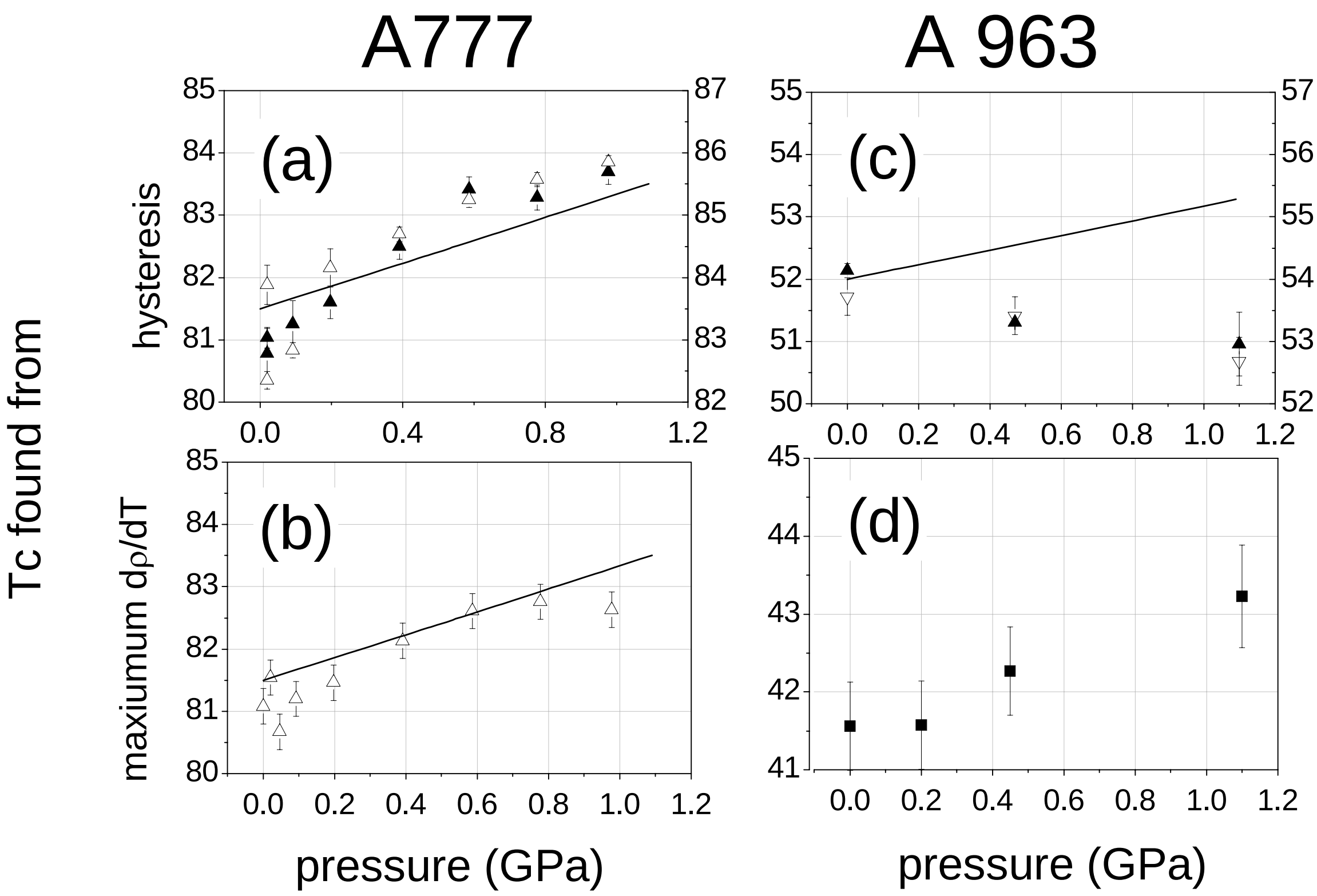

FiG. 7. Pressure dependence of Curie temperature deduced from the hysteresis loops detected by Hall resistivity – (a), (c), and from the maximum of *dρ/dT* – (b), (d). Figures (a) and (b) refer to the sample A777 and (c) and (d) – to the sample A963. The points give the values averaged over different pressure cycles. The two sets of points visible in Hall resistivity results (a), (c) correspond to different Hall probes. Although there is a difference in the absolute value of the $T_C$, the observed pressure trends are identical. The solid lines show the predictions of the model, described in the text, assuming that the free hole concentration changes due to volume changes only.